\documentclass[12pt,preprint]{aastex}
\usepackage{times}

\newcommand{\myemail}{iab\@roe.ac.uk}

\slugcomment{}

\shorttitle{OGLE~2003--BLG--235/MOA~2003--BLG--53}
\shortauthors{Bond et al.}

\begin{document}


\title{OGLE~2003--BLG--235/MOA~2003--BLG--53: A planetary microlensing event.}


\author{
I.A. Bond\altaffilmark{1}, 
A. Udalski\altaffilmark{2}, 
M. Jaroszy\'nski\altaffilmark{2,4},
N.J. Rattenbury\altaffilmark{3}, 
B. Paczy\'nski\altaffilmark{4}, 
I. Soszy\'nski\altaffilmark{2},
L. Wyrzykowski\altaffilmark{2}, 
M.K. Szyma\'nski\altaffilmark{2}, 
M. Kubiak\altaffilmark{2}, 
O. Szewczyk\altaffilmark{2,4}, 
K. \.Zebru\'n\altaffilmark{2}, 
G. Pietrzy\'nski\altaffilmark{2,5},  
F. Abe\altaffilmark{6}, 
D.P. Bennett\altaffilmark{7}, 
S. Eguchi\altaffilmark{6}, 
Y. Furuta\altaffilmark{6}, 
J.B. Hearnshaw\altaffilmark{8}, 
K. Kamiya\altaffilmark{6}, 
P.M. Kilmartin\altaffilmark{8}, 
Y. Kurata\altaffilmark{6}, 
K. Masuda\altaffilmark{6}, 
Y. Matsubara\altaffilmark{6}, 
Y. Muraki\altaffilmark{6}, 
S. Noda\altaffilmark{9}, 
K. Okajima\altaffilmark{6}, 
T. Sako\altaffilmark{6}, 
T. Sekiguchi\altaffilmark{6},  
D.J. Sullivan\altaffilmark{10}, 
T. Sumi\altaffilmark{4}, 
P.J. Tristram\altaffilmark{3},
T. Yanagisawa\altaffilmark{11}, and P.C.M. Yock\altaffilmark{3}
\\(The MOA and OGLE Collaborations)
}

\altaffiltext{1}{Institute for Astronomy, University of Edinburgh, Edinburgh, EH9 3HJ, UK;
iab@roe.ac.uk}
\altaffiltext{2}{Warsaw University Observatory, Al. Ujazdowskie 4, 00-478 Warszawa, Poland;
\{udalski, mj, soszynsk, wyrzykow, msz, mk, szewczyk, zebrun, pietrzyn\}@astrouw.edu.pl}
\altaffiltext{3}{Department of Physics, University of Auckland, Auckland, New Zealand;
nrat001@phy.auckland.ac.nz, p.yock@auckland.ac.nz, paulonieka@hotmail.com}
\altaffiltext{4}{Department of Astrophysical Sciences, Princeton University, Princeton NJ 08544, USA;
\{bp, sumi\}@astro.princeton.edu}
\altaffiltext{5}{Universidad de Concepcion, Departmento de Fisica, Casilla 160-C, Concepcion, Chile}
\altaffiltext{6}{Solar-Terrestrial Environment Laboratory, Nagoya University, Nagoya 464-8601, Japan;
\{abe, sada, furuta, kkamiya, kurata, kmasuda, ymatsu, muraki, sako, sekiguchi\}@stelab.nagoya-u.ac.jp}
\altaffiltext{7}{Department of Physics, Notre Dame University, Notre Dame, IN 46556, USA;
bennett@emu.phys.nd.edu}
\altaffiltext{8}{Department of Physics and Astronomy, University of Canterbury, Private Bag 4800, 
Christchurch, New Zealand; \{john.hearnshaw, pam.kilmartin\}@canterbury.ac.nz}
\altaffiltext{9}{National Astronomical Observatory of Japan, Tokyo, Japan; sachi.t.noda@nao.ac.jp}
\altaffiltext{10}{School of Chemical and Physical Sciences, Victoria University, PO Box 600, Wellington, 
New Zealand; denis.sullivan@vuw.ac.nz}
\altaffiltext{11}{National Aerospace Laboratory, Tokyo, Japan; tyanagi@nal.go.jp}


\begin{abstract}
We present observations of the unusual microlensing event 
OGLE~2003--BLG--235/MOA~2003--BLG--53. In this event a short
duration ($\sim$7 days) low amplitude deviation in the light
curve due a single lens profile was observed in both the MOA
and OGLE survey observations. We find that the observed features of the 
light curve can only be reproduced using a binary microlensing model
with an extreme (planetary) mass ratio of $0.0039^{+11}_{-07}$ for the
lensing system. If the lens system comprises a main sequence
primary, we infer that the secondary is a planet of
about 1.5 Jupiter masses with an orbital radius of $\sim3$ AU. 
\end{abstract}





\section{Introduction}

Gravitational microlensing occurs when a foreground object passes
through or  very near the line of sight of a background source star
generating a well known symmetric light curve profile.
If the foreground  lens object is a star with an orbiting
planet, then the presence of the planet  may be detectable via a brief
disturbance in the single lens light curve \citep{mp, gl}. This effect
can potentially be utilized to detect  planets with masses ranging from
those of gas giants right down to terrestrial planets \citep{br}.

The short timescales of these deviations, ranging from a few days for
giant  planets to hours for terrestrial planets, and their
unpredictability, present  considerable challenges in any observational
program. While some encouraging results have been obtained
\citep{benn99, alb00, rhie, bond02, jp}, no
firm detections of planets by microlensing have previously been
obtained.

In this Letter we report observations, obtained by OGLE and
MOA, of the event OGLE~2003--BLG--235/MOA~2003--BLG--53 (hereafter O235/M53)
that was independently detected in both survey programs. We observed a 7 day
deviation that was strongly detected in both surveys. We show that an
extreme mass ratio binary microlensing model best reproduces the
observed features in the light curve.

\section{Observations}

Presently, Galactic bulge microlensing events are discovered and then
alerted  by the two independently operating survey groups OGLE
\citep{udal03} and MOA \citep{bond01}. The microlensing event
OGLE~2003--BLG--235 ($\alpha$=18:05:16.35, $\delta$=$-$28:53:42.0, J2000.0)
was first identified and alerted by the OGLE EWS system \citep{udal03}
on 2003 June 22. It was independently detected by MOA on 2003 July 21
and alerted as MOA~2003--BLG--53.

OGLE observations were carried out with the 1.3 m Warsaw telescope at Las
Campanas Observatory, Chile, which is operated by the Carnegie Institute
of Washington, equipped with a mosaic CCD camera with 8192$\times$8192
pixels. The images were obtained in the I band with an exposure time of
120 seconds each. The observations presented here come from the OGLE-III 
phase of the OGLE survey and started in August 2001. Additional photometry 
of the
star was also collected during OGLE-II phase (1997--2000). This dataset
indicates, however, no variability of the object during that period and
was not used in further analysis.

MOA observations were carried out from the Mt John Observatory in New
Zealand with a 0.6 m telescope equipped with a mosaic CCD camera with  
4096$\times$6144 pixels. The MOA images were obtained using 180 second
exposures with a broad band red filter with its throughput centered on 
the standard I band.

The photometry was derived using difference imaging analysis carried 
out independently by the OGLE and MOA teams on their respective
datasets. This method is the current state-of-the-art for photometric
accuracy in crowded fields \citep{alard, alcock99}. Our analysis 
resulted in two sets of time-series photometry
in the I band corresponding to 183 OGLE measurements and 1092 MOA
measurements during 2000--2003. 

In Fig.~\ref{fig:lc} we show the light curve for this event on various
timescales from March 2000 to the present. The long term behavior is 
typical  for single point mass microlensing events, and it is similar
to the almost 2000 other events discovered in the Galactic 
Bulge since 1993. The unique feature of O235/M53 is, however, 
a short duration deviation from the profile expected for a single
lens, seen clearly in both datasets during 2003 July 14--21. Moreover,
a spike, characteristic of those binary microlensing events where the 
source enters or exits a bounded ``caustic'' region in the magnification
map projected on the sky, was observed and well covered by MOA on 
2003 July 21. This caustic region was crossed in 12\% of the overall
lens Einstein radius crossing time. This short duration, combined with
the small ($\sim$25\%) amplitude of the photometric deviation in the 
caustic region interior, suggests an extreme mass ratio binary system.

As well as regular monitoring in the I band, several V band
observations were obtained by OGLE at various magnifications of the
event. These were not used in the microlensing
modeling, but they were used to constrain the source and lens star
properties. By plotting the linearized fluxes in the I and V bands
against each other, a
model independent measurement of the color index of the source star was
determined. We obtained $\mathrm{V-I}=1.58\pm0.02$. Using 
$\mathrm{E(V-I)}=0.82$ mag for the interstellar reddening towards the source 
\citep{sumi}, the corrected color index of $(\mathrm{V-I})_0=0.76\pm0.02$ 
indicates a G type source star.

\section{Light curve modeling}

\begin{figure}
\epsscale{1.0}
\plotone{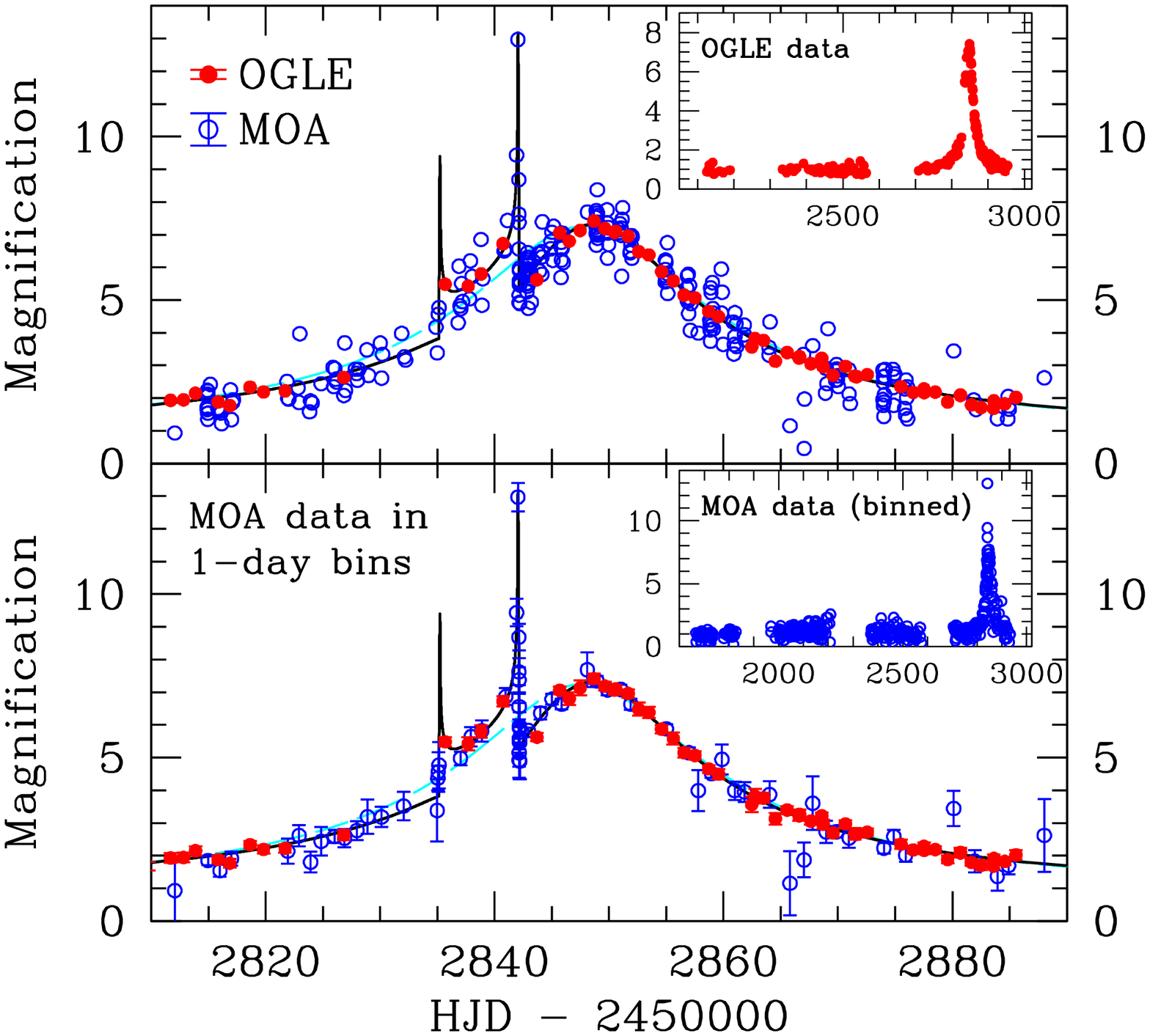}
\caption{Light curve with best fitting and single lens models of
O235/M53. The OGLE and MOA measurements are shown as red filled circles
and open blue circles, respectively. The top panel presents the complete
dataset during 2003 (main panel) and 2001--2003 OGLE data (inset). For
clarity, the error bars were not plotted, but the median errors in the
OGLE and MOA points legend are indicated. The lower panel is as the top
panel but with the MOA data grouped in 1 day bins except for the
caustic crossing nights, and with the inset showing MOA photometry
during 2000--2003. The binary and single lens fits are plotted in flux units normalized to
the unlensed source star brightness of the best planetary fit, and they
are indicated by the solid black and cyan dashed curves, respectively.
\label{fig:lc}}
\end{figure}

The modeling of the observed light curve of O235/M53 was 
performed independently by three groups using different methods to
generate numerical binary microlensing light curves \citep{br, ml,
ratt}, and all three found the solution that is 
presented in Fig.~\ref{fig:lc}. The observable quantities for all 
microlensing events are the Einstein radius crossing time,
$t_\mathrm{E}$, the impact parameter $u_0$ (in Einstein radius units)
of the source
star trajectory with respect to the lens center-of-mass (c.o.m.), and the time,
$t_0$, of the closest approach to the c.o.m. For binary microlensing 
events, one also measures the mass ratio, $q \equiv M_1/M_2$, the 
transverse separation, $a$, of the lens components, and the position
angle, $\phi$, of the binary with respect to the source--lens 
transverse velocity. For
caustic crossing events, one also measures the ratio, 
$\rho \equiv \theta_*/\theta_\mathrm{E}$, of the apparent angular
radius of the source star to that of the Einstein ring. 
In addition to these 7 physical parameters, there are two linear
scaling parameters between the 
magnification and the flux units for each passband, giving a total of 
11 parameters for the modeling. In our modeling procedure, we
searched for local $\chi^2$ minima
using minimization procedures that allowed all 11 parameters to vary 
simultaneously. Our light curve modeling also employed a surface limb
darkening profile appropriate for a G type star assuming a
metalicity that is approximately solar.

In Table~\ref{tab:param}, we list the physical microlensing parameters for the
best fitting model shown in Fig.~\ref{fig:lc}. This model has
$\chi^2=1390.49$ for 1267 degrees of freedom and an extreme mass ratio
of $q=0.0039$, which is a strong indication that the secondary may be
a planet. Since microlensing light curves generally allow a much more 
accurate determination of the secondary:primary mass ratio, than the
absolute mass of either body, the most sensible way to distinguish
planetary microlensing events from those due to binary systems is
through a criterion based on the mass ratio parameter $q$. 
There is a well known minimum in the distribution of mass ratios for
binary stars and planetary systems, which is known as the ``brown dwarf
desert\rlap." There are few systems known with $0.01 \lesssim q \lesssim 0.1$
\citep{hamuq,chab,mz}. Thus it
is sensible to define the boundary between stellar binary and
planetary binary microlensing events in between these $q$ values to
minimize any possible ambiguity. This leads to a criterion of $q<0.03$
for a planetary microlensing event, and so O235/M53 is clearly in the
planetary event category.

\begin{figure}
\epsscale{1.0}
\plotone{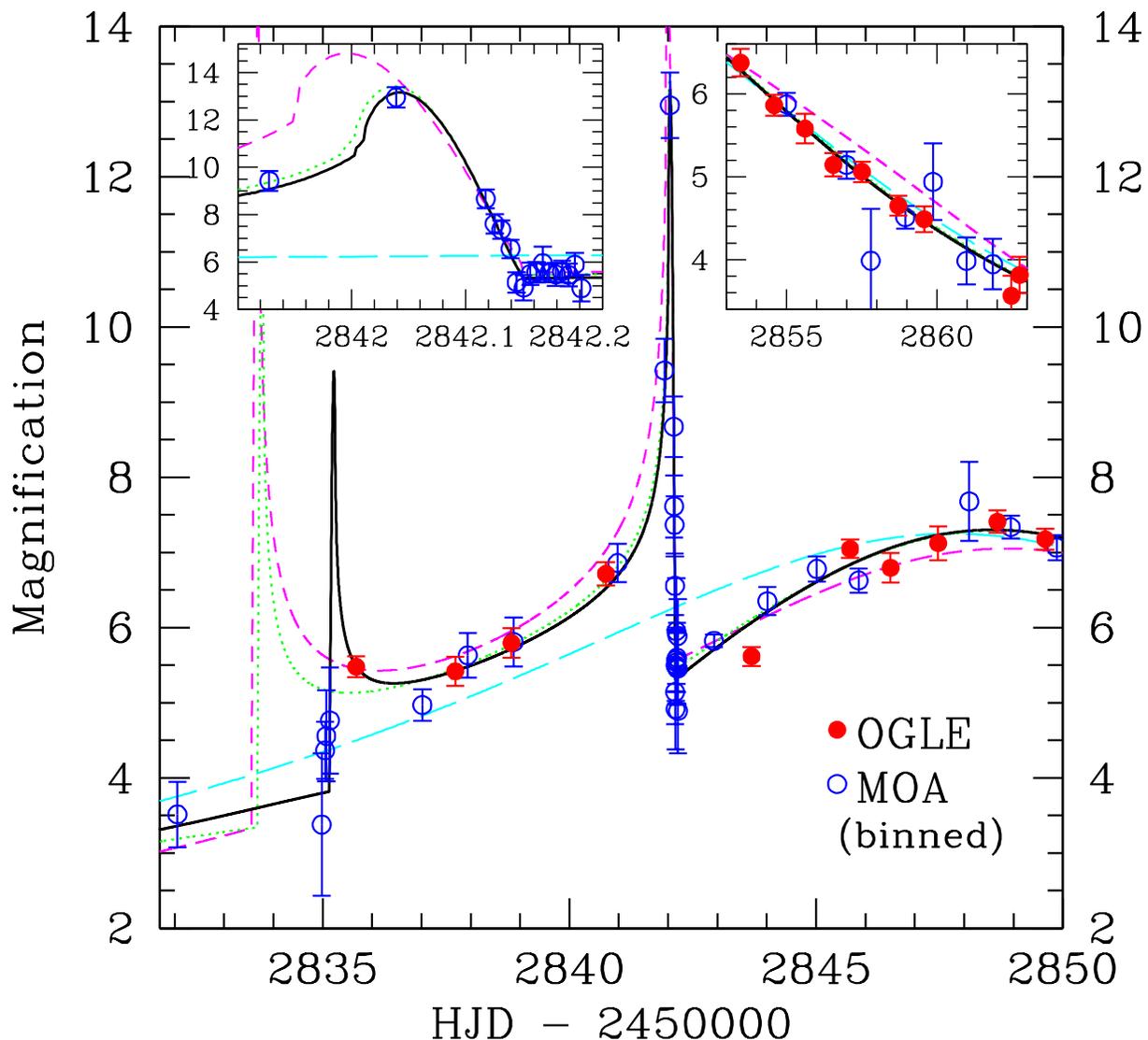}
\caption{Light curve and modeling of O235/M53
during the caustic traverse. The models shown are the best fits for
various regimes of the lens system parameter space, normalized as in 
Fig.~\ref{fig:lc}. These models are:
single lens case (cyan, long dash curve), binary lens with $q\ge 0.03$
(magenta, short dash line), planetary lens with caustic entry before day
2835 (green, dotted line), and the best overall fit with $q = 0.0039$
(dark, solid line). The insets show the second caustic crossing and 
a region of the declining part of the light curve where the best fit 
non-planetary binary lens model fails to fit the data. MOA data on days other
than the caustic entry and exit (days $2835\pm0.5$ and $2842\pm0.5$) are
placed in one day bins.
\label{fig:caustic}}
\end{figure}

We have carried out a systematic search in parameter space to try to
find sets of model parameters that might explain the observed light
curve with a larger mass ratio. Binary microlensing models with 
$q\ge 0.1$ that traverse a caustic curve in $\sim7$ days 
have much larger magnifications inside the caustic curve than
is observed for O235/M53. These binary lens events also have much
larger deviations from a single lens light curve before and after the
caustic crossings. As the mass ratio is decreased, the best fit light
curves approach the observed light curves, with much weaker caustic
crossing deviations. In Fig.~\ref{fig:caustic},
we show a close-up of the 7 day deviation with the best fit planetary
model, compared with the best non-planetary model with $q\ge 0.03$ and
best fit single lens model. The non-planetary binary models and single
lens models are strongly disfavored with fit $\chi^2$ values that are
larger by $\Delta\chi^2=210.96$ and $\Delta\chi^2=650.96$
respectively. In both cases, the $\chi^2$ improvement for the best fit
model is quite significant in both the MOA and OGLE data sets (see
Table~\ref{tab:param}). The failure of the non-planetary binary model
can be seen in the Fig.~\ref{fig:caustic} inserts.
This model predicts both stronger caustic signals and significant deviations
13--20 days after the second caustic crossing that are not consistent with
the observations. This model also shows some discrepancies at magnifications
$< 2$, but these are not as strongly excluded due to the higher photometric
uncertainties at lower magnification.

Also shown in Fig.~\ref{fig:caustic} is a
planetary model with an earlier caustic crossing and a larger planet
mass ratio: $q=0.0070$. This fit represents a distinct local minimum of
the $\chi^2$ surface, and is disfavored by only $\Delta\chi^2=7.37$
or $\sim2.7\sigma$. This is not accounted for by our 1--$\sigma$
uncertainty on $q=0.0039$. Therefore, we have increased the upper error
estimate on the planetary mass ratio to 0.0011, so that the actual
uncertainty in $q$ will be bounded by our error estimates at the
3--$\sigma$ level. In Table~\ref{tab:param} we also list the
parameters for these alternative models.

Finally, the first OGLE observation after the second caustic crossing
indicates a magnification below all of the binary models shown by about
$3.6\sigma$. Such an outlier is not unusual because the real photometric
error distributions for crowded field photometry generally have
larger wings than a Gaussian distribution. Three of the 183 OGLE measurements
are outliers from the best fit by $> 3\sigma$, and 16 are outliers by 
$> 2\sigma$. These outlier
points do not appear to cluster in the vicinity of the planetary deviation.
If this single data point did indicate a real light curve deviation, it
could be explained by a small variation to the planetary microlensing model,
such as a moon orbiting the planet, but there is no non-planetary model 
that could help to explain it.

\section{Further constraints on the source and lens}

Most microlensing events have only a single measureable parameter,
$t_\mathrm{E}$, that constrains the lens mass, distance, and
tranverse velocity with respect to the line-of-sight to the source.
However, time resolved observations of binary event caustic crossings 
resolve finite source star effects and partly remove these 
degeneracies \citep{alcock00, witt, nemiroff, gould94}, by allowing a 
measurement of the Einstein angular radius given by
$ \theta_\mathrm{E}^2 = ( 4GM_\mathrm{lens}/{c^2} )
   (D_\mathrm{source}-D_\mathrm{lens})/(D_\mathrm{source}D_\mathrm{lens})$.

Using the flux parameters of the microlensing fit, we obtained
$\mathrm{I}=19.70\pm0.15$ for the source star and $\mathrm{I}=20.7\pm0.4$
for the blended component. 
This source star magnitude, plus the V--I color from Section 2, can be
compared to the bulge color magnitude diagram of \citet{holtz}, and this
indicates that the source is probably a bulge star near the main sequence
turn-off.
To determine the angular radius of the source star
we used the color-color relations of \citet{bb} together with empirical
relations between $\mathrm{V-K}$ and surface brightness derived from
interferometry observations of nearby main sequence stars \citep{vanb, dib}.

We find $\theta_*=0.50\pm0.05$ $\mu$as, which combined with our measurement
of $\rho$, yields $\theta_\mathrm{E}=520\pm80$ $\mu$as. This yields
the following relation between the lens mass and distance
\begin{equation}
\frac{M_\mathrm{lens}}{M_\sun} = 0.123 \left( \frac{\theta_\mathrm{E}}{\mathrm{mas}} \right)^2
\left ( \frac{D_\mathrm{source}}{\mathrm{kpc}} \right) \frac{x}{1-x}
\label{eqn:md}
\end{equation}
where $x=D_\mathrm{lens}/D_\mathrm{source}$. If we combine this relation
with the mass luminosity relations of \citet{kt} for main sequence stars,
and require that the lens luminosity at a given distance does not exceed
the blend flux, we obtain an upper limit (90\% confidence) of 
$D_\mathrm{lens}<5.4$ kpc. Thus,
if the lens is a main sequence star, it must be in the Galactic disk.

\begin{figure}
\epsscale{.80}
\plotone{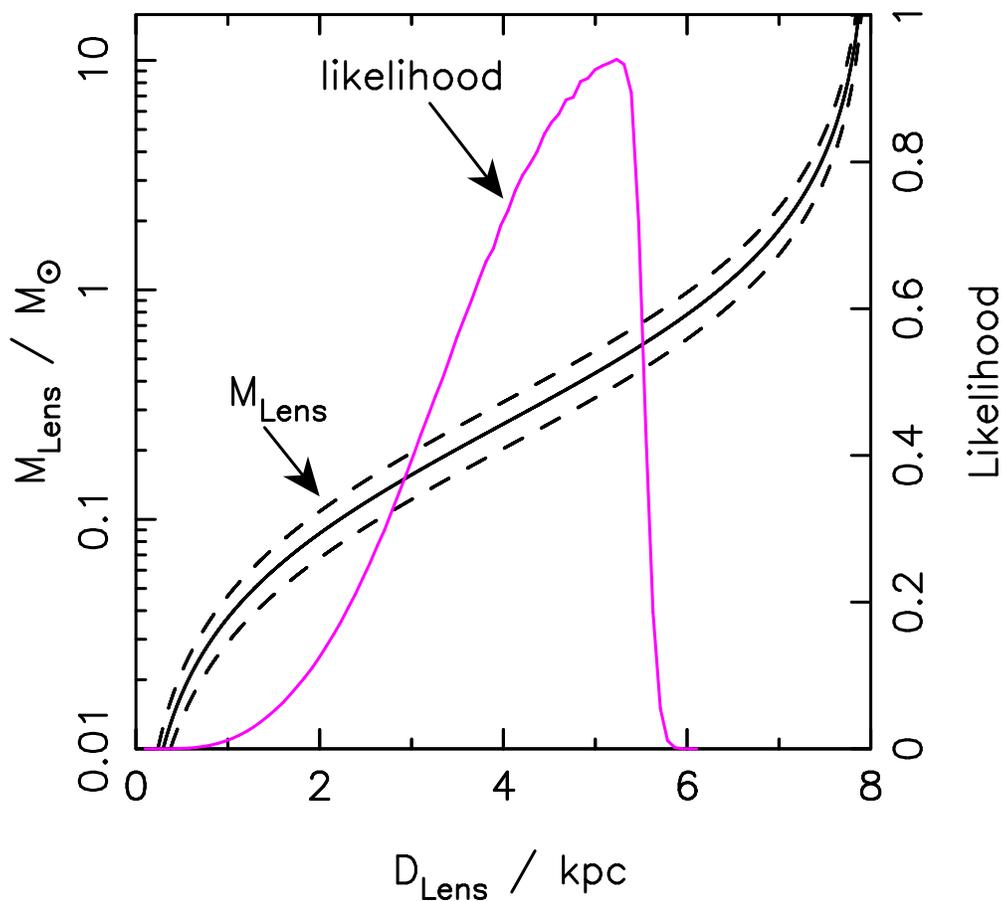}
\caption{
Constraints on the distance to O235/M53 and its mass.
Our measurement of the
angular size of the Einstein ring constrains the lens mass and distance
to lie on the solid curve. The curve is bounded above and below by the dark
dashed curves, due to the uncertainty in this measurement. The likelihood
function (shown by the magenta curve) is for a main sequence disk star prior
for the lens star with the sharp cut-off being due to the constraints 
imposed by the measured lens flux.
\label{fig:like}}
\end{figure}

In Fig.~\ref{fig:like}, we show Eqn.~\ref{eqn:md} together with the
results of a maximum likelihood analysis based on our measurements
of the Einstein ring and its characteristic crossing time. The likelihood
function was calculated using the Galactic disk models of \citet{hg}. We
then obtain with 90\% confidence: $D_\mathrm{lens}=5.2^{+0.2}_{-2.9}$ kpc from
which we infer the lensing system to comprise an M2--M7 dwarf star of 
mass $0.36^{+0.03}_{-0.28}$ M$_\sun$ with a giant planetary companion of 
$1.5^{+0.1}_{-1.2}$ M$_\mathrm{J}$ (Jupiter masses). 
The planet is in a wide orbit with a transverse separation of 
$3.0^{+0.1}_{-1.7}$ AU.

Another possibility for the lens is that it could be a remnant object
such as a white dwarf, neutron star, or black hole. If the lens is a
white dwarf with mass 0.6 M$_\sun$, Eqn.~\ref{eqn:md} would place it a
distance of 5.5 kpc. In this case, the microlensing parameters would
imply a 2.5 M$_\mathrm{J}$ planet orbiting the white dwarf with a
transverse separation of 2.8 AU.

\section{Discussion}

In Section~3 we concluded that the observed light curve of
O235/M53 is best described by a binary
lensing model with an planetary mass ratio of $q=0.0039$.
Our definition of the planetary nature of the secondary lens
by means of the mass ratio is optimal when the mass
ratio can be measured, but it is useful to consider other possible definitions.
Another potential dividing line between planets and brown dwarfs is 
the Solar metalicity threshold for sustained Deuterium burning at 
13.6 M$_\mathrm{J}$, although 
Deuterium burning itself has little relevance for planet formation.

The situation in the case of O235/M53 was helped
by the measurement of finite source effects. If the lens is a 
main sequence star, then as shown in the previous section, it must
be an M dwarf with a $\sim 1.5\,$M$_\mathrm{J}$ planetary
companion\footnote{The only 
other M dwarf star known to have planetary 
companions is Gliese 876 \citep{marcy}.}. 
There is a non-negligible chance that the lens is a
white dwarf and a much smaller chance that it is a neutron star, but
in both cases, a planetary companion below the nominal 13.6 M$_\mathrm{J}$
threshold is required. Only in the unlikely case of a massive black hole
primary, could the secondary be outside the range traditionally
associated with a planet.

There are some prospects for follow-up observations of this event. Our
measurements of the finite source effects imply a proper motion of
the lens with respect to the source of 
$\mu = \theta_\mathrm{E}/t_\mathrm{E} = 3.1 \pm 0.4~\mathrm{mas/yr}$.
High resolution imaging carried out $\sim 10\,$years from now with JWST or
adaptive optics systems should be able to resolve the lens and source stars
providing direct measurements \citep{han,alcock01} of the color and
brightness of the lens, as well as confirmation of the proper motion
measurement.

We present these observations as a demonstration of the planetary microlensing
phenomenon.  The power of microlensing is in its
ability to acquire statistics on many systems \citep{gest}. 
These include planets in wide orbits, very low mass planets, and even
planets in other galaxies \citep{covone, bond02}. 
The challenge now to the microlensing
community is to develop effective strategies to find more planetary 
microlensing
events. 

Numerical photometry of OGLE~2003--BLG--235/MOA~2003--BLG--53 is 
available from the websites for OGLE \url{http://ogle.astrouw.edu.pl} 
and MOA \url{http://www.physics.auckland.ac.nz/moa}.

\acknowledgments
The MOA project is supported by the Marsden Fund of New Zealand, the 
Ministry of 
Education, Culture, Sports, Science, and Technology (MEXT) of Japan, and the 
Japan Society for the Promotion of Science (JSPS). Partial support to the 
OGLE project was provided by the following grants: the Polish State Committee 
for Scientific Research grant 2P03D02124 to A. Udalski and 2P03D01624 to 
M. Jaroszy\'nski, the NSF grant AST-0204908 and NASA grant NAG5-12212 to 
B. Paczy\'nski. A.U., I.S., and K.Z. also acknowledge support from the 
grant "Subsydium Profesorskie" of the Foundation for Polish Science. 
Support was also provided by grants NSF AST-0206187 and NASA NAG5-13042 
to D. Bennett.

\clearpage



\clearpage
\begin{deluxetable}{llllllllllll}
\rotate
\tablecaption{Microlens Model Parameters\label{tab:param}}
\tablewidth{0pt}
\tabletypesize{\scriptsize}
\tablehead{
  \colhead{Model} &
  \colhead{$M_\mathrm{p}/M_\star$} &
  \colhead{$\theta_*/\theta_\mathrm{E}$} &
  \colhead{$a_\mathrm{proj}/R_\mathrm{E}$} &
  \colhead{$\phi$} &
  \colhead{$u_0$} &
  \colhead{$t_0$} &
  \colhead{$t_{\rm E}$} &
  \colhead{$I_{\rm source}$} &
  \colhead{$\chi^2$} &
  \colhead{$\chi^2_{\rm MOA}$} &
  \colhead{$\chi^2_{\rm OGLE}$} \\
  \colhead{} &
  \colhead{} &
  \colhead{} &
  \colhead{} &
  \colhead{} &
  \colhead{} &
  \colhead{} &
  \colhead{(days)} &
  \colhead{mag} &
  \colhead{(1267 dof)} &
  \colhead{(1089 dof)} &
  \colhead{(178 dof)}
}
\startdata
Best Fit & $0.0039({+11\atop -07})$ & $0.00096(11)$ &
        $1.120(7)$ & $223\fdg8(1\fdg4)$ & $0.133(3)$ &
        $2848.06(13)$ & $61.5(1.8)$ & 19.70(15)& $1390.49$ & $1151.00$ & $239.50$ \\
Early Caustic & $0.0070$ & $0.00104$ & $1.121$ & $218\fdg9$ & $0.140$ &
        $2847.90$ & $58.5$ & 19.62 & $1397.87$ & $1149.37$ & $248.49$ \\
Best Non-planet & $0.0300$ & $0.00088$ & $1.090$ & $187\fdg9$ & $0.144$ &
        $2846.20$ & $57.5$ & 19.68 & $1601.44$ & $1229.47$ & $371.98$ \\
Single Lens & -- & -- & -- & -- & $0.222$ &
        $2847.77$ & $45.2$ & 19.10 & $2041.45$ & $1624.17$ & $417.28$ \cr
\enddata
\tablecomments {
The units for $t_0$ are ${\rm HJD} - 2450000$.
}
\end{deluxetable}

\end{document}